\documentstyle[preprint,aps,prd]{revtex}

\begin{document}

\title{
Dynamics of Collapse of a Confined Bose Gas}
 
\author{L.P. Pitaevskii}
 \address{Department of Physics, Technion, 32000 Haifa, Israel\\ and\\ 
Kapitza Institute for Physical Problems, 117454 Moscow, Russia}
 
 \date{April 11, 1996}
 
 \maketitle
 
 \begin{abstract}
Rigorous results on the nonlinear dynamics of a dilute Bose gas 
with a negative scattering length in an 
harmonic magnetic trap are presented and sufficient conditions for
the collapse of the system are formulated. By using the virial theorem for
the Gross-Pitaevskii equation in an external field we analyze the 
temporal evolution of the mean square radius $I=\int r^2\mid 
\psi\mid^2dV$ of the gas cloud. In the 2D case this quantity 
undergoes harmonic 
oscillations with frequency $2\omega_0$. It implies that for 
a negative value of the
energy of the system, the gas cloud will collapse after a finite time interval.
For positive energy the cloud collapses if the initial conditions 
correspond to a large
 enough amplitude of oscillations. Stable oscillations with a small amplitude
are possible. In the 3D case the system also 
collapses after a finite time for a state with negative 
energy. A stringent condition for the collapse is also derived.
 \end{abstract}
 

\narrowtext
The recent observation of Bose-Einstein condensation of atoms of $^{87}$Rb
 \cite{1},
$^{7}$Li \cite{2} and $^{23}$Na \cite{3} 
confined in magnetic traps has opened a new important 
field of investigation of quantum 
phenomena. In particular, a new feature of 
these systems is that they are significantly 
inhomogeneous so one can directly observe effects
 of quantum uncertainty on a macroscopical scale.
 The formal evidence of such a situation is that dynamics of these 
non-uniform gases at low temperatures can be properly described by a 
non-linear equation
for the condensate wave function $\psi({\bf r},t )$ \cite{4}, \cite{5}:

 \begin{eqnarray}
 i\hbar\frac{\partial}{\partial t} \psi({\bf r},t) =
 \left (-\frac{\hbar^2\nabla^2}{2m} + U({\bf r})+g
\mid \psi({\bf r},t)\mid^2 \right )\psi({\bf r},t) .
 \label{1}
 \end{eqnarray}
Here $U$ is the 
confining potential and 
$g=\frac{4\pi {\hbar}^2 a}{m}$,
 where $a$ 
 is the 
s-wave scattering length. This equation is quite classical in its merit
 but contains explicitly the Planck's constant $\hbar$ . (We will put
$\hbar = m = 1$ below.)

Experimental evidences of BEC in magnetically trapped alkali
vapour stimulated recent numerical investigations
 of Eq. (\ref{1}) in static, linear and nonlinear regimes. \cite{6} - \cite{10}

The physical situation is very different for $^{87}$Rb, $^{23}$Na 
on one side and $^7$Li on the other. In the first case $a>0$ and 
the homogeneous state of the gas is stable. For $^7$Li 
the scattering length $a$
is negative ($a=-1.43 \times 10^-7$ cm), 
which corresponds to an effective attraction, and the gas 
possesses
negative compressibility. The uniform state in this case is unstable and
the gas can undergo a "collapse" which is an unrestricted contraction.
(The idea of such a collapse was first  introduced by V.Zacharov
for a formally similar system of Langmiur waves \cite{11}.) Fetter 
however has recently shown \cite{7}, using
a variational approach, first employed by Baym and Pethick 
\cite{6}, that a system of a finite number $N$ of atoms
in a trap can be in a metastable state for $N$ below some critical number 
$N_c$ . The same conclusion has been obtained by Ruprecht et al.
\cite{8}. Surprisingly the number of atoms reached in the experiment \cite{2}
appears significantly larger than the theoretical estimate  of $N_c$ .

In such an uncertain situation it is important to understand better the
conditions and dynamics of the collapse.

In this paper we will present some rigorous results relative to the behaviour
of the solution of Eq. (\ref{1}). Particularly we will 
formulate a sufficient 
condition for collapse in the gas. We will analyze the temporal evolution 
of the quantity $I=\int r^2 \mid \psi \mid ^2 dV$, which is proportional 
to the average square of the radius of the gas cloud. 
This "virial theorem"
has been used by Vlasov et al. to analyze the problem of the collapse,
described by Eq.(\ref{1}), in the absence of the confined field \cite{12}.
(See also more recent papers \cite{13} -\cite{16} and references therein.)

Note first of all that Eq. (\ref{1}) provides conservation laws for the
energy $E$ of the system:
\begin{eqnarray}
E=\int \left (\frac {1}{2} \mid \nabla \psi \mid ^2 + \frac {g}{2}\mid 
\psi \mid^4 + U\mid \psi \mid^2 \right )dV \equiv E_0 + \int U \mid \psi 
\mid^2 dV \label{2}
\end{eqnarray}
and for the number of atoms $N=\int \mid \psi \mid ^2dV$. The last 
conservation law is expressed by the continuity equation:
\begin{eqnarray}
\frac{\partial \mid \psi \mid ^2}{\partial t} + \frac{\partial
j_{i}}{\partial x_i}=0 .
\label{3}
\end{eqnarray}
The equation for the momentum density in the presence of an external field
can be written in the form:
\begin{eqnarray}
\frac{\partial j_i}{\partial t}+\frac{\partial \Pi _{ik}}{\partial x_k}
=-\mid \psi \mid ^2 \frac{\partial U}{\partial x_i} .
\label{4}
\end{eqnarray}

We will not write explicitly the well known expressions for the currents
$j_i$ and $\Pi _{ik}$ and we note only that in 2D and 3D we have an exact 
identities
\begin{eqnarray}
\int \Pi _{ii}dV=2E_0
\label{5}
\end{eqnarray}
and
\begin{eqnarray}
\int \Pi _{ii} dV=
3E_0-\frac{1}{2}\int \mid \nabla  \psi \mid ^2 dV  
\label{6}
\end{eqnarray}
respectively.

Repeating differentiation of $I$ with respect to $t$ and integrating by parts
one gets:
\begin{eqnarray}
\frac{dI}{dt}=2\int x_ij_idV, \frac{d^2I}{dt^2}=2\int \Pi _{ii}dV-2\int x_i
\frac{\partial U}{\partial x_i}\mid \psi \mid ^2 dV.
\label{7}
\end{eqnarray}

We first analyze the two-dimensional motion of the gas in an isotropic 
cylindrical 
trap with a confining potential $U=\frac{\omega _0^2r^2}{2}, r^2=x^2+y^2$.
 In this case Eq. (\ref{7}) takes a surprisingly simple form:
\begin{eqnarray}
\frac{d^2I}{dt^2}=-4\omega _0^2I + 4E.
\label{8}
\end{eqnarray}
whose general solution is of the obvious form:
\begin{eqnarray}
I=A\cos(2\omega _0t+\gamma)+E/\omega _0^2 .
\label{9}
\end{eqnarray}
It is worth noting that the parameters $A,\alpha$ and $E$ of the solution 
(\ref{9}) can be calculated using the initial distribution $\psi ({\bf r},0)$ 
with the help of Eq.(\ref{2}) and the first equality in Eq.(\ref{7}).

	Eq. (\ref{8}) describes harmonic pulsations of $I$ with 
constant frequency $2\omega_0$ in spite of the nonlinear character of  
Eq.(\ref{1}). (It is possible to show that small axial
symmetric oscillations of 
the system have the frequency $2\omega _0$  \cite{17}.) Furthermore, one 
can see that for a negative value 
of the energy the positive quantity $I$ approaches zero after a finite time 
smaller
than $\pi/2\omega _0$. This means complete contraction of the cloud. On the 
contrary, for a positive value of the energy
pulsations with a small enough amplitude $A<E/\omega _0^2$ do not result 
in a collapse \cite{18}. (See also the discussion in the concluding part of 
the paper.) 

Thus increasing the energy improves the 
stability of the cloud. For example an
increase in the stability has been noted in
\cite{19} in the presence of a vortex line.

Equation (\ref{8}) has the same form both for 
attractive and repulsive interactions. In the latter 
case however the collapse does not occur. The confining potential energy 
$\omega _0^2I/2$ is then always smaller than $E$. This inequality ensures 
that the collapse cannot be reached. 

One can suggest that under the experimental conditions  
some motion of the
superfluid gas exists that increases the energy of the system. This means 
that 
the cloud is not in an equilibrium but rather in a nonstationary stable state.

Results for the 3D case are not so conclusive. Consider an anisotropic
harmonical trap with a confining potential of the form 
$U=\frac{\alpha _{ik}x_ix_k}{2}$. One gets from (\ref{7}) and (\ref{6}):
\begin{eqnarray}
\frac{d^2I}{dt^2}=6E-5\int \alpha _{ik}x_ix_k dV - 2K,
\label{10}
\end{eqnarray}
where $K=\frac{1}{2} \int \mid \nabla \psi \mid ^2 dV$ is the kinetic energy.
 Eq. (\ref{10}) gives the 
inequality: 
\begin{eqnarray}
\frac{d^2I}{dt^2}+5\omega _m^2I-6E < 0,
\label{11}
\end{eqnarray}
where $\omega _m^2$ is the smallest eigenvalue of the $\alpha _{ik}$ matrix.
We can prove again that the system collapses after a finite time 
for a negative $E$.
Consider the auxiliary equation
\begin{eqnarray}
\frac{d^2I}{dt^2}+5\omega _m^2I -6E =0 .
\label{12}
\end{eqnarray}
For negative $E$ the quantity $I$ reaches, according to this equation, the value
zero after a time smaller than $\frac{\pi}{\sqrt{5}\omega _m}$. The negative
term in the left hand side of Eq.(\ref{10}) accelerates this 
process for the same initial 
conditions. The same situation takes place for a positive energy for a large
enough amplitude. 

One can get more rigorous sufficient conditions of the collapse using the 
powerful method developed in \cite{13}-\cite{15}. In this method an 
important part is played by two unequalities imposed on the kinetic energy
$K$\cite{13}, \cite{14} . The first one is the uncertainty relation between 
$I$ and $K$: 
\begin{eqnarray}
IK \geq \frac{9}{8}N^2 .
\label{13}
\end{eqnarray}
The second inequality can be written as
\begin{eqnarray}
\frac{1}{2} \int \mid \psi \mid ^4 dV \leq 
\beta N^{1/2}K^{3/2}, \beta = 0.0575.
\label{14}
\end{eqnarray}
Combining (\ref{13}) and (\ref{14}) one immediately gets an inequality 
for the energy $E$ as a function of $K$:
\begin{eqnarray}
E(K) > K+\frac{9}{16}\omega _m ^2 N^2 K^{-1} - \beta \mid g \mid
K^{3/2} N^{1/2}.
\label{15}
\end{eqnarray}
This equation demonstrates a remarkable resemblance to the expression for 
$E(d^{-2})$ which has been obtained by Fetter using a wave function of the 
form $\psi = C\exp (-r^2/2d^2)$. (See Eq.(\ref{5}) in \cite{7}.) For such 
a function the kinetic energy is proportional to $d^{-2}$. It easily checked
that $E(d^{-2})$ (expressed as a function of $K$) is larger that the right 
hand side of (\ref{15}).
The equation (\ref{15}) permits us to formulate the stringent collapse 
conditions we are looking for.
Introducing dimensionless variables $\epsilon$ and $\kappa$ defined by
\begin{eqnarray}
E = \frac{3}{4}N\omega _m \epsilon, 
K = \frac{3}{4}N\omega _m \kappa, 
\label{16}
\end{eqnarray}
we can rewrite this equation as:
\begin{eqnarray}
\epsilon > \kappa + \kappa ^{-1} - 1.18 \frac{2}{3}\sigma 
\equiv f(\kappa, \sigma ),
\label{17}
\end{eqnarray}
where $\sigma $ is a dimensionless parameter
\begin{eqnarray}
\sigma  = \sqrt{2/\pi}N\mid a \mid \omega _m ^{1/2}
\label{18}
\end{eqnarray}
The parameter $\sigma $ is analogous to the parameter $\sigma$ defined 
in \cite{7}. 

Behaviour of the function $f(\kappa)$ depends on the value 
of $\sigma $. For $\sigma  \geq \sigma _c = 0.454$ the curve is a 
monotonous one (the curve (a) in Fig.1). 
Let the point $\epsilon ^*$ , $\kappa ^* $ to be an intersection of the curve
$\epsilon = f(\kappa)$ and the straight line $\epsilon = \kappa /3$. Let 
us to be
\begin{eqnarray}
E < E^* = \frac{3}{4}N\omega _m \epsilon ^* >0.
\label{19}
\end{eqnarray} 
Then the kinetic energy $K  > K ^* = 3E^*$ for any $t$. This means that 
equation (\ref{10}) leads to the inequality
\begin{eqnarray}
\frac{d^2I}{dt^2}+5\omega _m^2I -6(E-E^*) < 0,
\label{20}
\end{eqnarray}
This formula obviously demonstrates that at condition (\ref{19}) we get
the collapse.

We shall now discuss the case where $\sigma  < \sigma _c$. The curve 
$\epsilon= f(\kappa)$ has in this case a minimum $\epsilon _m$ and a maximum
$\epsilon _M$ (the curve (b) in Fig.1).
Let us consider an initial state, where
\begin{eqnarray}
\epsilon _m <\epsilon < \epsilon _M, \kappa _1 < 
\kappa < \kappa _2. 
\label{21}
\end{eqnarray} 
Since any solution of (\ref{1}) evolves continuously, the kinetic energy
$\kappa$ will stay in the same range at every time. The uncertainty relation
(\ref{13}) gives then
\begin{eqnarray}
I>\frac{3N}{2\omega _m \kappa _2 }.
\label{22}
\end{eqnarray}
Thus the mean square radius stays restricted and collapse of the cloud as 
a whole is impossible.

If $E<E^*$ the cloud will collapse as in the previous case. The evolution 
of the cloud for $E>E^*$ and $K$ out the interval (\ref{21}) 
demands additional investigation. 

The method we explored is also useful in the two-dimensional case. In 
this case instead of (\ref{13}) and (\ref{14}) we have
\begin{eqnarray}
IK \geq \frac{1}{2}N^2 ,\frac{1}{2} \int \mid \psi \mid ^4 dV \leq
\beta NK, \beta = 0.17.
\label{23}
\end{eqnarray} 
We introduce dimensionless variables 
$\epsilon$ and $\kappa$ defined by:
\begin{eqnarray}
E = \frac{1}{2}N\omega _m \epsilon,
K = \frac{1}{2}N\omega _m \kappa .
\label{24}
\end{eqnarray}
Then we obtain the following inequality for the energy: 
\begin{eqnarray}
\epsilon > \kappa (1-N/N_c) + \kappa ^{-1} 
\equiv g(\kappa, N ),
\label{25}
\end{eqnarray}
where $N_c=1/\beta \mid g \mid$
The shape of the right hand side of this inequality is different for $N
> N_c$ and for $N<N_c$ . 
(See curves (a) and (b) in Fig.2 respectively.)
In the latter case function
$g$ tends to $+ \infty$ as $\kappa$ tends both to 0 or $\infty$. Hence 
the kinetic energy for a given value $E$ stays restricted by some values
$\kappa _1$ and $\kappa _2$ and the total collapse is impossible. This 
means that for $N<N_c$ an arbitrary initial function $\psi ({\bf r},0)$
automatically ensures such initial conditions for Eq. (\ref{8}) that the 
collapse to be absent. In the opposite case, when  $N>N_c$, it is 
impossible to draw by this method any general conclusions.

I would like to thank S. Stringari for various enlightening discussions 
and S. Giorgini for useful remarks. I am grateful for hospitality to the
Department of Physics of the University of Trento where this paper has 
been completed.\\

\begin{figure}
\caption{ The function $f(\kappa )$ defined in (17)  for $\sigma =
0.51> \sigma _c$ (curve (a))  and for $\sigma =0.29< \sigma _c$ (curve (b)).
Admited states of the system are above the bold curve.
The dashed line is $\epsilon = \kappa /3$.}
\caption{ The function $g(\kappa )$ defined in (25)  for $N =
0.4N_c $ (curve (a))  and for $N =
1.4N_c $ (curve 
(b)). Admited states of the system are above the corresponding curve.}

\end{figure} 
  

\begin{thebibliography} {99}
\bibitem{1} M.N. Anderson, J.R. Ensher, M.R. Matthews, 
C.E. Wieman, and E.A. Cornell, Science {\bf 269}, 198 (1965).
\bibitem{2} C.C. Bradley, C.A. Sacket, J.J. Tollet, and R.G. Hulet, Phys. 
Rev. Lett. {\bf 75}, 1687 (1995). 
\bibitem{3}  K.B. Davis, M.-O. Mewes, M.R. Andrews, N.J. van Druten, 
D.S. Durfee,
D.M. Kurn, and W. Ketterly, Phys. Rev. Lett. {\bf 75}, 3969 (1995).
\bibitem{4} E.P. Gross, Nuovo Cimento {\bf 20}, 454 (1961); J. Math. Phys.
 {\bf 46} 137 (1963).
\bibitem{5} L.P. Pitaevskii, Sov. Phys.  JETP {\bf 13}, 451 (1961).
\bibitem{6} G. Baym and C. Pethick, Phys. Rev. Lett. {\bf 76}, 6 (1996).
\bibitem{7} A. Fetter, (unpublished).
\bibitem{8} P.A. Ruprecht. M.J. Holland, K. Burnett, and M. Edwards, Phys. 
Rev. A {\bf 51}, 4704 (1995).
\bibitem{9} P.A. Ruprecht, M. Edwards, and K. Burnett, (unpublished).
\bibitem{10} M. Holland and J. Cooper, Phys. Rev. A {\bf 53}, R1954 (1996).
\bibitem{11} V.E. Zakharov, Sov. Phys. JETP {\bf 35}, 908 (1972).
\bibitem{12} S.N. Vlasov, V.A. Petrishcev, and V.I. Talanov,
 Izv. Vyssh. Uchebn. Zaved., Radiofiz. {\bf 12}, 1353 (1970).
\bibitem{13} M.I. Weinstein, Commun. Math. Phys. {\bf 87}, 567 (1983).
\bibitem{13a} E.A. Kuznetzov, A.M. Rubenchik, and V.E. Zakharov, 
Phys. Rep., {\bf 142}, 103 (1986). 
\bibitem{14} S.K. Turitsyn, Phys. Rev. E {\bf 47}, R13 (1993).
\bibitem{15} E.A. Kuznetsov, J.J. Rasmussen, K. Rypdal, and S.K. 
Turitsun, Physica D {\bf 87}, 273 (1995).
\bibitem{16} P.M. Lushnikov, JETP Lett., {\bf 62}, 461 
(1995). \bibitem{17} S. Stringari, (private communication).
\bibitem{18} One must be careful about interpretation of these
oscillations. The present considerations cannot exclude 
the possibility of a
"partial collapse" of  the denser part of the cloud.
Collapsed atoms will escape the trap and the system will fit
itself to a more stable state.
The occurence of a
singular point where the density formally tends to infinity is also 
possible. 
\bibitem{19} F. Dalfovo and S. Stringari, Phys. Rev. A. {\bf 53}, (1996).
 
 \end{thebibliography}
\end{document}